\documentclass[conference]{IEEEtran}
\IEEEoverridecommandlockouts
\usepackage[a4paper,left=0.625in, right=0.625in, top=0.7in,bottom=1.07in]{geometry}
\usepackage[utf8]{inputenc}
\usepackage{cite}
\usepackage{amsmath,amssymb,amsfonts}
\usepackage{algorithmic}
\usepackage{graphicx}
\usepackage{textcomp}
\usepackage{url}
\usepackage[caption=false]{subfig}
\usepackage[justification=centering]{caption}
\usepackage[ruled,vlined]{algorithm2e}
\usepackage{epsfig}
\usepackage{float}
\usepackage{color}
\usepackage{balance}
\usepackage{bbm}
\usepackage{textcomp}
\usepackage{enumitem}
\usepackage{mathtools}
\def\BibTeX{\text{B\kern-.05em{\sc i\kern-.025em b}\kern-.08em
    T\kern-.1667em\lower.7ex\hbox{E}\kern-.125emX}}

\newcommand{\beq}{\begin{equation}}
\newcommand{\eeq}{\end{equation}}

\floatstyle{ruled}
\newfloat{algorithm}{tbp}{loa}
\providecommand{\algorithmname}{Algorithm}
\floatname{algorithm}{\protect\algorithmname}

\def\BibTeX{{\rm B\kern-.05em{\sc i\kern-.025em b}\kern-.08em
    T\kern-.1667em\lower.7ex\hbox{E}\kern-.125emX}}
\begin{document}

\title{Energy-Efficient Classification at the\\ Wireless Edge with Reliability Guarantees\vspace{-.2 cm}
\thanks{This work has been partly funded by the European Commission through the H2020 project Hexa-X (Grant Agreement no. 101015956).}}

\author{Mattia Merluzzi$^1$, Claudio Battiloro$^2$, Paolo Di Lorenzo$^2$, Emilio Calvanese Strinati$^1$\\
$^1$CEA-Leti, Université Grenoble Alpes, F-38000 Grenoble, France\\
$^2$Dept. of Information Engineering, Electronics, and Telecommunications, Sapienza University of Rome, Italy.\\
email:\{mattia.merluzzi, emilio.calvanese-strinati\}@cea.fr,\{claudio.battiloro,paolo.dilorenzo\}@uniroma1.it\vspace{-.2 cm}}

\maketitle

\begin{abstract}
Learning at the edge is a challenging task from several perspectives, since data must be collected by end devices (e.g. sensors), possibly pre-processed (e.g. data compression), and finally processed remotely to output the result of training and/or inference phases. This involves heterogeneous resources, such as radio, computing and learning related parameters. In this context, we propose an algorithm that dynamically selects data encoding scheme, local computing resources, uplink radio parameters, and remote computing resources, to perform a classification task with the minimum average end devices' energy consumption, under E2E delay and inference reliability constraints. Our method does not assume any prior knowledge of the statistics of time varying context parameters, while it only requires the solution of low complexity per-slot deterministic optimization problems, based on instantaneous observations of these parameters and that of properly defined state variables. Numerical results on convolutional neural network based image classification illustrate the effectiveness of our method in striking the best trade-off between energy, delay and inference reliability.
\end{abstract}


\section{Introduction}
While 5G networks are already in their deployment phase, the race to 6G has just started, and the path is still long towards a first standardization. One of the fundamental challenges of 6G is to design a \textit{holistic} system in which communication, computation, and control are jointly orchestrated to achieve new target levels of dependability, energy efficiency, and sustainability. In this ecosystem, a central role is played by Machine learning (ML) and Artificial Intelligence (AI), and especially by their deployment at the edge of wireless networks, close to the end consumers, towards the rising concept of Edge ML/Edge AI, or more in general Edge Intelligence (EI) \cite{Zhou19_EI}. EI comes with a twofold perspective: i) \textit{EI for communications}, i.e. ML/AI based methods enabling wireless networks' optimization with unprecedented flexibility; ii) \textit{Communications for EI}, i.e. the exploitation of massive amounts of connections to collect, transmit and process enormous data volumes within low end-to-end (E2E) delays, possibly with a sustainable perspective. The aforementioned perspectives are also naturally bonded into an end-to-end view in which AI/ML, and in general adaptive mechanisms, are exploited to manage radio, computing and control resources, to finally enable reliable and sustainable AI/ML on data collected at the edge \cite{MerluzziEML2021}.
\begin{figure*}[t]
    \centering
    \includegraphics[width=.79\textwidth]{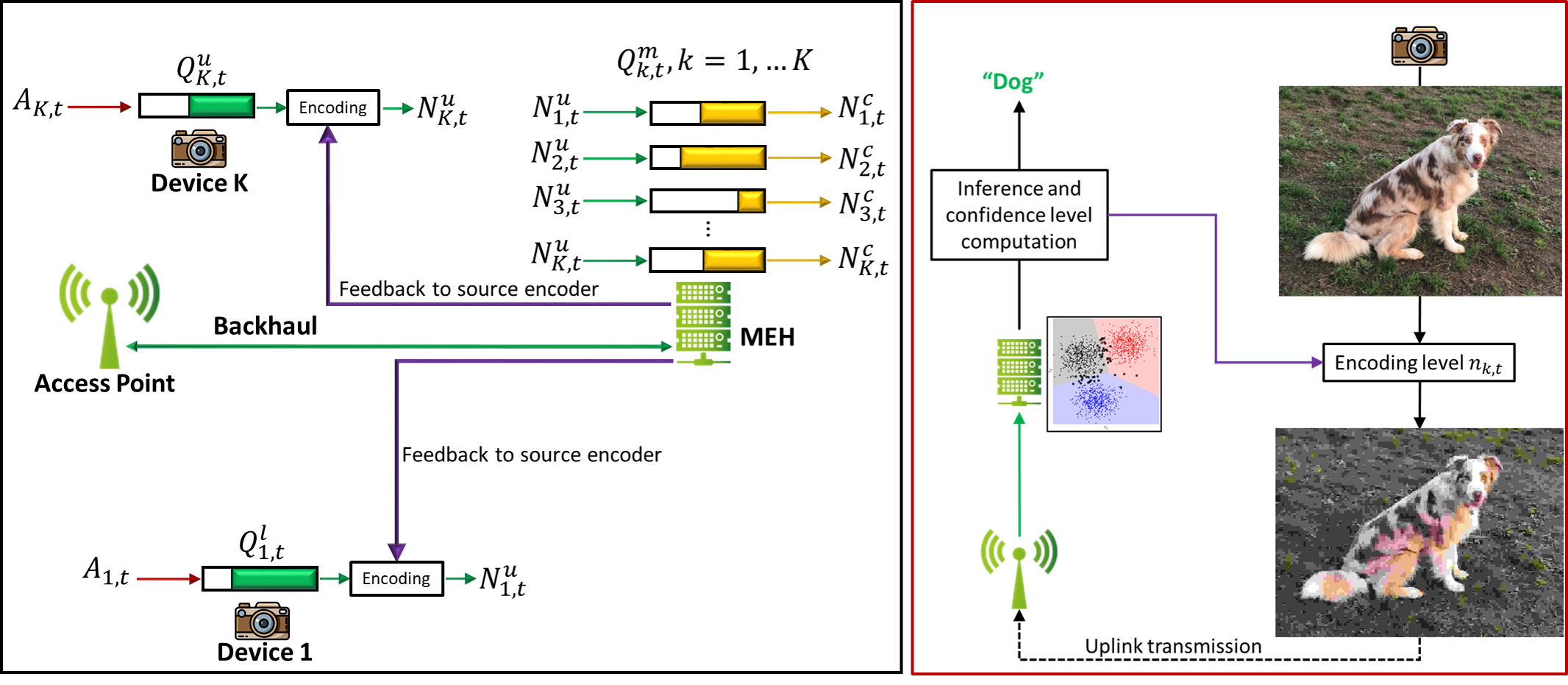}
    \caption{Network model}
    \label{fig:scenario}
    \vspace{-.4 cm}
\end{figure*}
A key enabler of EI is Multi-access Edge Computing (MEC) \cite{ETSIMEC}, which foresees the deployment of distributed computing resources close to end users, namely in Mobile Edge Hosts (MEHs) that are, e.g. co-located with radio Access Points (APs). In this work, we focus on the paradigm of communications for EI, devising a resource allocation framework to enable dynamic energy efficient edge classification on data collected by end devices, with target E2E delays and inference reliability constraints. \\
\textbf{Related works.} As pointed out in \cite{Zhou19_EI,Letaief22}, several aspects must be tackled when performing edge inference tasks in the most efficient way, such as model compression techniques to reduce the memory footprint \cite{Han15}, model selection strategies to adaptively select the inference model online \cite{Tay18}, or application oriented optimization, for instance adapting frame rate and resolution of video streaming for resource efficient classification \cite{Jia18}, \cite{Ran18}. The authors of \cite{Gal21} aim at maximizing the average accuracy of a video analytics task, under a frame rate and delay constraint. Also, \cite{LiYong17} explores the trade-off between energy, delay and accuracy, by optimizing a parameter that controls the quality of the results and the workload size. Similarly, in \cite{LiuHuang18}, the authors formulate the edge inference problem as the minimization of a weighted sum of latency and a decreasing function of the analytics accuracy, with a per device analytics accuracy constraint. In \cite{MerluzziEML2021}, a joint management of radio and computation resources is proposed to explore the trade-off between energy, latency and accuracy, exploiting data quantization as a way to control the accuracy, for different learning tasks. 
\\
\textbf{Contribution.} The aim of this work is to perform another step towards a holistic system, in which communication, computing and learning aspects are jointly optimized in system design and operation phases. We propose an adaptive resource allocation algorithm aimed at enabling energy efficient image classification with E2E delay and inference reliability guarantees. We consider a dynamic scenario, in which data (images) are continuously collected by end devices (e.g. cameras), and uploaded to an MEH through the wireless connection with an AP. 
To cope with time varying context parameters, such as radio channels and data arrivals, our method \textit{dynamically, adaptively, and jointly} optimizes; i) Data encoding scheme; ii) Radio resources; iii) Local and remote computation resources. We propose to use a specific metric of inference reliability, whose dynamic behaviour is measured through formally defined state variables that can be updated online, and controlled by the encoding scheme with a closed-loop. Numerical results on a Convolutional Neural Network (CNN) based image classification assess the performance of our strategy in striking the best trade-off between energy consumption, E2E delay, and inference reliability, which directly affects the final accuracy. 

\section{System model and performance indicators}\label{sec:system_model}
In this work, we focus on edge inference, assuming a pre-trained and pre-uploaded model at the edge (i.e. at the MEH). We consider a dynamic system, with time-varying context parameters (i.e. wireless channels and data arrivals). In such a dynamic setting, data are continuously generated locally at each device (e.g. a camera capturing images), and uploaded to an MEH that runs an inference task (e.g. object recognition). Since we deal with image classification, we design a procedure in which data are: i) Collected and buffered locally at the device; ii) Encoded and transmitted; iii) Remotely buffered and processed by the MEH for classification. While the exploited queueing system is represented on the left hand side of Fig. \ref{fig:scenario}, we present a high level description of the edge inference process through the flow diagram on the right hand side. After being generated by the device, data are encoded (with an optimized compression level), depending on the channel conditions, queues' states, etc., as we will describe later on. This results into a new (possibly degraded) version of each pattern, which is then transmitted to the MEH through the wireless connection with the AP (uplink transmission). At this point, the MEH can process data to output the inference result (i.e. a label) and, a the same time, a confidence level associated with its decision (e.g. the entropy, as it will be clarified later on), which represents our metric of inference reliability. The latter is used to control the source encoder, to close the loop between encoding, communication, and computing.
The goal of this procedure is to provide inference results within a finite E2E delay, with the least device energy consumption and target inference reliability.\\
\textit{Notation:} The operator $\lfloor\cdot\rfloor$ represents the floor operation, i.e. the highest integer lower than the argument; $\mathbf{1}\{\cdot\}$ is the indicator function. Also, let us denote the long-term average of a generic random variable $X$ as
\begin{equation}\label{average_value}
    \overline{X} = \lim_{T\to\infty}\frac{1}{T}\sum\nolimits_{t=1}^T\mathbb{E}\left\{X_t\right\},
\end{equation}
where the expectation is generally taken with respect to random context parameters (e.g. time-varying wireless channels and data arrivals). The above notation ($\overline{X}$) is useful to represent the long-term average of all variables used throughout the paper. All other notation details will be given in the text, when needed.
\subsection{End-to-end delay}
In our setting, edge inference entails four phases: i) Local buffering; ii) Encoding and transmission; iii) Remote buffering; iv) Remote computation for inferencing. We choose to neglect the time needed to send the results back to the devices, which in this case is only a scalar denoting the label, i.e. a very small amount of information. We consider a time-slotted dynamic system, where each slot is indexed by $t$ and has duration $\tau$. To characterize the E2E delay, let us consider a 2-hop queueing system, with a local communication queue at each device and a remote computation queue at the MEH for each device, whose time evolution is described in the following. 
\subsubsection{Local buffering, encoding and transmission}
At time $t$, the $k$-th local communication queue is fed by new arrivals, denoted by $A_{k,t}$. At the same time, given an uplink transmit power $p_{k,t}^u$, the data rate can be written as follows:
\begin{equation}\label{uplink_rate}
    R_{k,t}^u=B_{k,t}\log_2\left(1+\frac{h_{k,t}^u p_{k,t}^u}{N_0B_{k,t}}\right),
\end{equation}
where $h_{k,t}^u$ is the  time varying channel power gain between device $k$ and the AP, $B_{k,t}$ is the bandwidth assigned to device $k$ at time $t$, and $N_0$ is the noise power spectral density at the receiver. We assume that patterns are encoded right before transmission, while they live in form of raw data until they soujour in the local communication queue. To this end, we assume that the whole slot duration $\tau$ is divided into two portions: a portion $\tau_e$ for encoding, and a portion $\tau_u$ for uplink transmission. We denote by $n_{k,t}$ the number of bits used to encode the patterns transmitted during time slot $t$. Also, let us denote by $f_{k,t}^l$ the local CPU cycle frequency used by device $k$ to perform the computations necessary to encode patterns, and by $J_k^e$ the number of CPU cycles needed to encode one pattern. Therefore, assuming that all encoded patterns should be transmitted (and vice versa) the number of transmitted patterns during time slot $t$ is
\begin{equation}\label{uplink_data}
    N_{k,t}^u=\left\lfloor \tau_u R_{k,t}^u/n_{k,t}\right\rfloor=\left\lfloor\tau_e f_{k,t}^l/J_{k}^e\right\rfloor,
\end{equation}
which is guaranteed by imposing the straightforward relation
\begin{equation}\label{change_var}
    f_{k,t}^l=\tau_u R_{k,t}^u J_k^e/(\tau_e n_{k,t}),
\end{equation}
which will be used later on in the problem formulation. As it will be clarified, the linear relation between $R_{k,t}^u$ and $f_{k,t}^l$ preserves important structures in the problem, which will help us deriving low complexity solutions, besides clearly linking radio and computation resources, with a single variable for each user (i.e., $R_{k,t}^u$). Finally, the local buffer evolves as:
\begin{align}\label{local_queue}
    Q_{k,t+1}^u=\max\left(0,Q_{k,t}^u-N_{k,t}^u\right)+A_{k,t}.
\end{align}
\subsubsection{Remote buffering and inferencing}
The remote buffer at the MEH is fed by the transmitted patterns, and drained by remote processing (i.e. by providing inference outputs). Specifically, denoting by $f_{k,t}^r$ the portion of the MEH CPU clock frequency dedicated to device $k$ (depending on the optimized MEH scheduling policy), and by $J_k^c$ the number of CPU cycles needed to classify one pattern, the number of patterns classified during time slot $t$ is
\begin{equation}\label{num_computed}
    N_{k,t}^c=\left\lfloor\tau f_{k,t}^r/J_k^c\right\rfloor.
\end{equation}
A remote queue is defined for each device, as different devices compete for a limited pool of computing resources, to be efficiently scheduled. The $k$-th queue evolves as follows: 
\begin{align}\label{comp_queue}
    &Q_{k,t+1}^r=\max\left(0,Q_{k,t}^r-N_{k,t}^c\right)+\min(N_{k,t}^u,Q_{k,t}^u),
\end{align}

\subsubsection{Stability and average E2E delay}\label{sec:delay}
The described 2-hops queueing system is propaedeutic to characterize the E2E delay of the inferencing process. In particular, denoting by $\overline{Q_k^u}$, $\overline{Q_k^r}$ the long-term average of the involved queues (cf. \eqref{average_value}), the stability of the system is formalized as follows: 
\begin{align}\label{stability}
    & \overline{Q_k^u}<\infty,\;\forall k;\quad\overline{Q_k^r}<\infty,\;\forall k.
\end{align}
Moreover, in a stationary stable system, due to Little's law \cite{Little1961}, the average E2E delay is finite and can be written as $\overline{D}_k=\tau(\overline{Q_k^u}+\overline{Q_k^r})/\overline{A_k}$,
where $\overline{A_k}$ is the average number of new arrivals $A_{k,t}$ (cf. \eqref{local_queue}), computed as in \eqref{average_value}.
\subsection{Energy consumption}
In this paper, we are interested in the end devices' energy consumption, while we leave the whole network energy-efficiency for future investigations. In the proposed scenario, devices spend energy to encode (i.e. for local processing) and to upload data through the uplink wireless transmission.
Recalling  the local CPU cycle frequency $f_{k,t}^l$ used by device $k$ at time $t$, the encoding energy consumption at time $t$ is:
\begin{equation}\label{encoding_energy}
    E_{k,t}^e=\tau_e\kappa_k \left(f_{k,t}^l\right)^3=\tau_e\kappa_k \left(\tau_u R_{k,t}^u J_k^e/(\tau_e n_{k,t})\right)^3,
\end{equation}
where $\kappa_k$ is the effective switched capacitance of the processor \cite{Burd1996}, and the second equality directly comes from \eqref{change_var}. At the same time, recalling the uplink transmit power $p_{k,t}^u$ of device $k$ at time $t$, the energy spent for transmission is
\begin{equation}\label{tx_energy}
    E_{k,t}^u=\tau_u p_{k,t}^u=\frac{\tau_u N_0B_{k,t}}{h_{k,t}^u}\left[\exp\left(\frac{R_{k,t}^u\ln(2)}{B_{k,t}}\right)-1\right],
\end{equation}
where the second equality comes from the inverse of \eqref{uplink_rate}.
\subsection{Inference reliability}\label{sec:inf_rel}
\begin{figure}[t]
    \centering
    \includegraphics[width=\columnwidth]{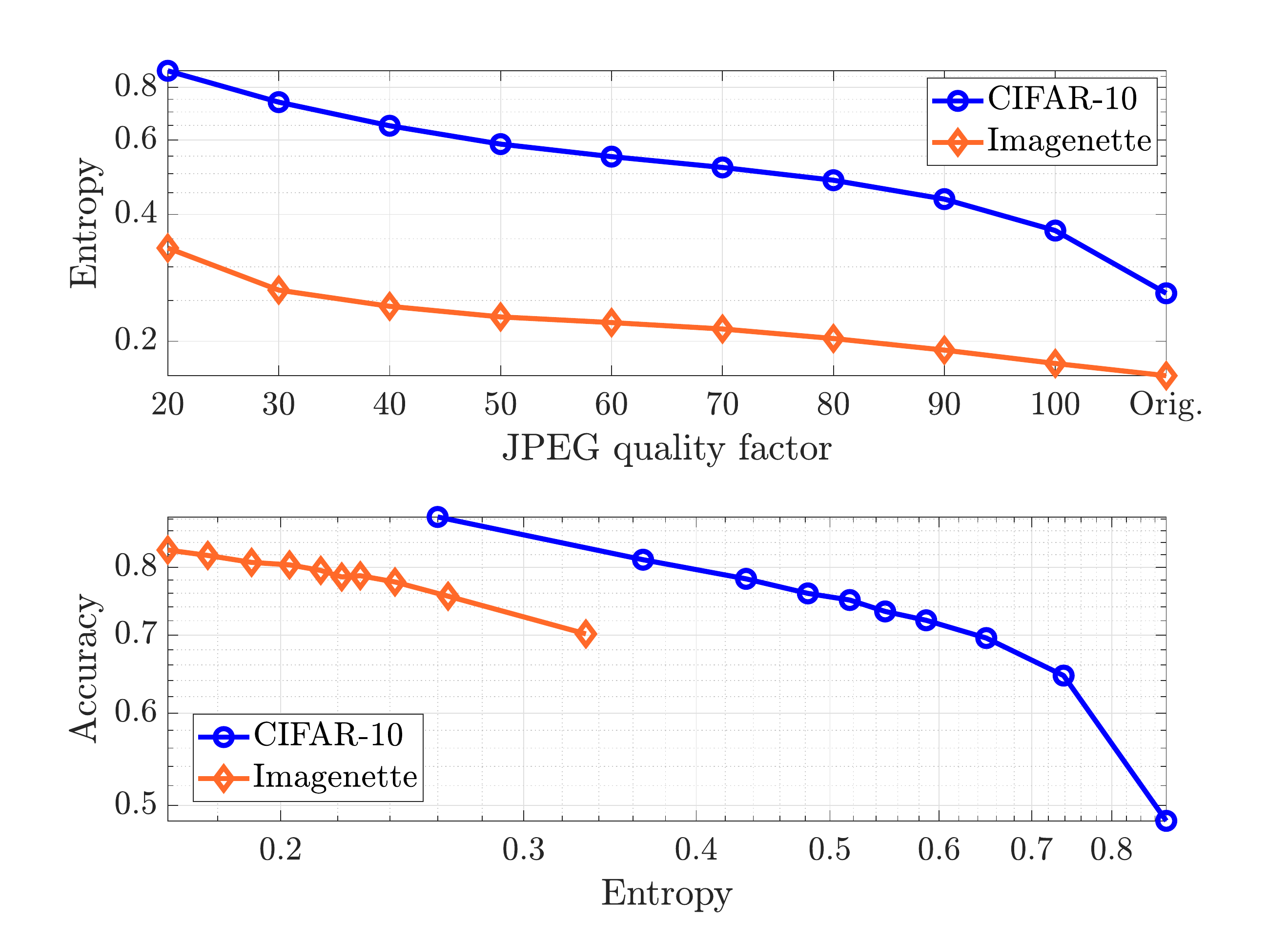}
    \caption{Accuracy and entropy vs. JPEG quality factor}
    \label{fig:acc_entr_jpeg}
    \vspace{ 0 cm}
\end{figure}
The last performance indicator we exploit is a measure of the inference reliability. It should be noted that a proper measure should not require any ground truth to be able to work online on previously unseen (i.e. unlabeled) data. Therefore, it is not possible to directly use the accuracy (i.e. correct classification rate) as a measure of inference reliability. In this paper, we consider the entropy computed on a posteriori probabilities assigned to each label, as a measure of confidence in classifying a given pattern. For instance, given a pattern $i$ generated by user $k$, and $l_k=1,\ldots L_k$ possible labels, the output of a neural network based classifier is a vector of probabilities $[p_{k,i,1},\ldots p_{k,i,L_k}]$. The entropy associated with this vector reads as $H_{k,i}=-\sum\nolimits_{l=1}^{L_k}p_{k,i,l}\log(p_{k,i,l}).$
The reasoning to use the entropy as inference reliability measure is threefold: i) It does not require any ground truth; ii) It applies generally to any discriminative or generative classifier \cite{Bishop06}; iii) The cross-entropy is the typical loss function used to train neural networks for classification. Finally, as reliability constraint, we choose to use the long-term average entropy, thus imposing the following constraint (cf. \eqref{average_value}):
\begin{equation}\label{entropy_constraint}
    \overline{H_{k,i}}\leq H_k^{\textrm{th}},\quad\forall k,
\end{equation}
where $H_k^{\textrm{th}}$ denotes an entropy threshold requested a priori by the intended device, and depends on the specific service and goal. Also, in this paper, we make the assumption that the output entropy is a function of the image quality. Indeed, we will consider the case of lossy compression, so that the inference task at the MEH is run on compressed images. 
Finally, the real metric (not measurable online) is the accuracy, which in this paper we mildly assume to be directly linked to the entropy. In particular, we make the following two assumptions to link data compression, entropy, and accuracy. \\ \textbf{\textit{Assumption 1:}} \textit{$H_{k,i}$ is a monotone non increasing function of the number of bits $n_{k,t}$ (cf. \eqref{uplink_data}) used to encode patterns.}\\
\textbf{\textit{Assumption 2:}} \textit{The accuracy (i.e. correct classification rate) is a monotone non increasing function of $H_{k,i}$.}\\
Assumptions $1$ and $2$ are not formally proved, but they are mild assumptions that we also validate through the results in Fig. \ref{fig:acc_entr_jpeg}, in which we show the the entropy as a function of the JPEG quality factor used to encode images, and the accuracy as a  function of the entropy, for two different data sets, namely CIFAR-$10$ \cite{krizhevsky2009learning} and Imagenette\footnote{\url{https://github.com/fastai/imagenette}}, as examples. Other examples are omitted due to the lack of space. 

\section{Problem formulation}
Recalling \eqref{average_value}, \eqref{change_var}, \eqref{stability}, \eqref{encoding_energy}, \eqref{tx_energy}, and \eqref{entropy_constraint}, we formulate the following long-term optimization problem:
\begin{align}\label{problem_edge_inference}
    &\hspace{-1.8 cm}\underset{\mathbf{\Psi}_t}{\min}\quad\sum\nolimits_{k=1}^K\left(\overline{E_k^e}+\overline{E_k^u}\right)\\
    \textrm{subject to}\quad&\eqref{stability},\;\eqref{entropy_constraint},\;(a)\: n_{k,t}\in\mathcal{N}_k\nonumber,\\
    &(b)\: I_{k,t}\widetilde{R}_{k,t}^{\min}(n_{k,t})\leq R_{k,t}^u\leq \widetilde{R}_{k,t}^{u,\max}I_{k,t},\; \forall k,t,\nonumber\\
    &(c)\: f_{k,t}^r\geq 0,\,\forall k,t,\;(d)\: \sum\nolimits_{k=1}^Kf_{k,t}^r\leq f_r^{\max},\,\forall t,\nonumber
\end{align}
where $\mathbf{\Psi}_t=[\{n_{k,t}\}_{k=1}^K,\{R_{k,t}^u\}_{k=1}^K,\{f_{k,t}^r\}_{k=1}^K]$, recalling that $f_{k,t}^l$ is directly linked to $R_{k,t}^u$ through \eqref{change_var}; whereas, $I_{k,t}=\mathbf{1}\{n_{k,t}\}$ equals $1$ if device $k$ encodes and transmits patterns during time slot $t$. Also, $\widetilde{R}_{k,t}^{\min}(n_{k,t})$ is defined as
\begin{equation}\label{rate_min}
\widetilde{R}_{k,t}^{\min}(n_{k,t})=\frac{n_{k,t}}{\tau_u},
\end{equation}
meaning that, if a user transmits, the data rate (and the local computation rate) must be as high as to transmit (and encode) at least one pattern. On the other hand, $\widetilde{R}_{k,t}^{u,\max}$ reads as
\begin{equation}\label{rate_max}
    \widetilde{R}_{k,t}^{\max}=\min\left(\frac{\tau_e n_{k,t}f_{k}^{l,\max}}{\tau_u J_k^e},B_{k,t}\log_2\left(1+\frac{h_{k,t}^uP_k^u}{N_0B_{k,t}}\right)\right),
\end{equation}
where $f_{k}^{l,\max}$ is the maximum local CPU cycle frequency, and $P_k^u$ is the maximum uplink transmit power.
Besides the long-term constraints already discussed in Section \ref{sec:system_model}, the instantaneous constraints in \eqref{problem_edge_inference} have the following meaning: $(a)$ The number of bits used to encode data (i.e. the encoding strategy) belongs to a discrete finite set $\mathcal{N}_k$, which includes $n_{k,t}=0$; $(b)$ If $n_{k,t}=0$, the user does neither encode nor transmit. Instead, if the user transmits (i.e. $n_{k,t}>0$), the transmit data rate is higher than the threshold defined in \eqref{rate_min} and lower than the maximum value defined in \eqref{rate_max};  $(c)$ The remote CPU cycle frequency assigned by the MEH to each user is non negative; $(d)$ The sum of the CPU frequencies assigned to each user is lower than the maximum CPU clock frequency of the MEH $f_r^{\max}$. 
Problem \eqref{problem_edge_inference} is very complex from several points of view. First of all, the expectations are taken with respect to wireless channels and data arrivals, whose statistics are supposed to be unknown a priori (i.e. the objective function is unknown). Moreover, it is non-convex and involves integer variables over a long-term horizon, which makes it a mixed integer non linear program. 
\subsection{The Lyapunov framework}\label{sec:Lyapunov}
Lyapunov stochastic optimization is a powerful tool that can be used to transform long-term problems into pure stability programs, which are then solved in a per-slot fashion, typically with low complexity or exhaustive searches over a reduced (instantaneous) action space \cite{Neely10}.  
The first goal is to define a mathematical tool able to deal with the long-term constraint \eqref{entropy_constraint}. To this end, we define a \textit{virtual queue} $Z_{k,t}$, which is updated each time a new pattern is classified, i.e. whenever a new measure of entropy is available. Therefore, its evolution between subsequent slots reads as follows:
\begin{equation}\label{Z_evolution}
    Z_{k,t+1}=\max\left(0, Z_{k,t}+\epsilon_z\sum_{i\in\mathcal{C}_{k,t}}\left(H_{k,i}-H_k^{\textrm{th}}\right)\right),
\end{equation}
where $\epsilon_z$ is a positive step size, and $\mathcal{C}_{k,t}$ is the set of all patterns classified during time slot $t$, whose cardinality is $N_{k,t}^c$ (cf. \eqref{num_computed}). Virtual queues generally measure the behaviour of a system in terms of constraint violations. Theoretically speaking, the \textit{mean rate stability}\footnote{It is defined as $\lim_{T\to\infty}\frac{\mathbb{E}\{Z_{k,T}\}}{T}=0$} of a virtual queue guarantees to meet the corresponding constraint \cite{Neely10}. This allows us to transform \eqref{problem_edge_inference} into a pure stability problem, in which we aim to guarantee the stability of both physical and virtual queues, to meet a finite average E2E delay and a target entropy, respectively. To this aim, defining the vector
 $\mathbf{\Theta}_t=[\{Q_{k,t}^u\}_{k=1}^K,\{Q_{k,t}^m\}_{k=1}^K,\{Z_{k,t}\}_{k=1}^K]$, we introduce the \textit{Lyapunov function} as follows \cite{Neely10}:
\begin{equation}\label{lyap_fun}
    L(\mathbf{\Theta}_t)=\frac{1}{2}\sum\nolimits_{k=1}^K\left[(Q_{k,t}^u)^2+(Q_{k,t}^r)^2+Z_{k,t}^2\right],
\end{equation}
which is a measure of the overall congestion state of the system in terms of both physical and virtual queues. Our aim is to push the network towards low congestion states (in terms of physical and virtual queues), while minimizing the energy consumption. To this end, let us define the \textit{drift-plus-penalty} (DPP) function $    \Delta_{p,t}=\mathbb{E}\left\{L(\mathbf{\Theta}_{t+1})-L(\mathbf{\Theta}_t)+VO_t\big|\mathbf{\Theta}_t\right\}$,
where $O_t=\sum\nolimits_{k=1}^KE_{k,t}^e+E_{k,t}^u$ (i.e. it is the instantaneous objective function of \eqref{problem_edge_inference}). The DPP is the conditional expected variation of the Lyapunov function over one slot, with a penalty factor, weighted by a parameter $V$, which assigns more or less priority to the objective function of \eqref{problem_edge_inference} (i.e. the devices' energy consumption) compared to queue backlogs, thus shaping the trade-off between energy consumption, E2E delay, and inference reliability. Interestingly, the stability of the physical and virtual queues is guaranteed if the DPP is bounded by a finite constant at each slot. Therefore, as in \cite{Neely10}, we proceed by minimizing an upper bound of the DPP, which in this case reads as
\begin{align}\label{DPP_upper}
  & \hspace{-.3 cm}\Delta_{p,t}\leq\zeta+\sum\nolimits_{k=1}^K\mathbb{E}\big\{(Q_{k,t}^r-Q_{k,t}^u)N_{k,t}^u-Q_{k,t}^r N_{k,t}^c\nonumber\\
    &\hspace{-.3 cm}+\epsilon_zZ_{k,t} \sum_{i\in\mathcal{C}_{k,t}}\left(H_{k,i}-H_k^{\textrm{th}}\right)+V\left(E_{k,t}^u+E_{k,t}^e\right)\big|\mathbf{\Theta}_t\big\},
\end{align}
where $\zeta$ is a positive constant omitted due to the lack of space, as well as the derivations leading to \eqref{DPP_upper}, which closely follow the arguments in \cite{Neely10}. 
As already mentioned in section \ref{sec:inf_rel}, the entropy is affected by the number of bits used to encode patterns. However, at time $t$ (when the decision on the number of bits $n_{k,t}$ is taken), the entropy value $H_{k,i}$ associated with number of bits $n_{k,t}$ is generally not available. Thus, to control efficiently the number of bits $n_{k,t}$ at time $t$, we introduce a surrogate function replacing $H_{k,i}$ in (\ref{DPP_upper}), say $\varphi(n_{k,t})$, which behaves similarly to $H_{k,i}$ with respect to $n_{k,t}$ (cf. Assumption 1). This falls under the concept of a $C$-additive approximation, for which the usage of outdated queue states does not prevent the solution to asymptotically approach optimal performance, with the cost of increased queues backlogs and convergence time \cite{Neely10}. Now, let us hinge on the concept of opportunistically minimizing \eqref{DPP_upper}, to optimize resources in a per slot fashion by removing the expectation. It can be easily shown that, in this case, the problem can be split into a radio resource allocation problem, involving local encoding and uplink transmission variables (i.e., $\{n_{k,t}\}_{k=1}^K$ and $\{R_{k,t}^u\}_{k=1}^K$), and a remote computation problem, to optimize the MEH CPU scheduling (i.e., $\{f_{k,t}^r\}_{k=1}^K$). The two problems, along with their respective solutions, are presented in the next two sections.
\subsection{Encoding and radio resource allocation sub-problem}
The sub-problem to be solved to encode and transmit patterns is a mixed-integer non linear program. In this work, we assume the bandwidth to be assigned a priori to each user. Therefore, the problem is separable among different users and can be formulated as follows
\begin{align}\label{radio_resource}
    \hspace{-.2 cm}\quad\underset{\{n_{k,t}\}_k,\{R_{k,t}^u\}_k}{\min} &O_{r,t}\coloneqq V(E_{k,t}^u+E_{k,t}^e)+(Q_{k,t}^r-Q_{k,t}^u)N_{k,t}^u\nonumber\\
    &\qquad+\epsilon_zZ_{k,t}\left(\varphi(n_{k,t})-H_k^{\textrm{th}}\right)N_{k,t}^u\\
    &\textrm{subject to}\quad (a)\text{-}(c)\;\text{of\;\eqref{problem_edge_inference}}\nonumber
\end{align}
Problem \eqref{radio_resource} is a mixed integer non linear program, although being much less complex than \eqref{problem_edge_inference}, since it requires only an exhaustive search over $\mathcal{N}_k$, whose cardinality is limited. Namely, for a generic user $k$, we can analyze two cases:\\
\textbf{Case 1:} $n_{k,t}=0$. In this case, the solution is $R_{k,t}^u=0$, and the value of the objective function is $O_{r,t}(n_{k,t}=0) = 0$.\\
\textbf{Case 2:} $n_{k,t}>0$. In this case, an exhaustive search over the small set $\mathcal{N}_k\setminus\{n_{k,t}=0\}$ must be performed. Interestingly, for each $n_{k,t}$, the optimal data rate can be found with extremely low complexity. In particular, by considering the tight approximation $\left\lfloor\frac{\tau_u R_{k,t}^u}{n_{k,t}}\right\rfloor\approx\frac{\tau_u R_{k,t}^u}{n_{k,t}}$, the problem is convex. 
Solving the Karush-Kuhn-Tucker (KKT) conditions \cite{boyd2004convex}, which are omitted due to the lack of space, it can be easily shown that, given $n_{k,t}$ the solution can be found by solving the following non linear equation:
\begin{align}\label{nonlineq}
    &3V\tau_e\kappa_k\left(\frac{\tau_uR_{k,t}^uJ_k^e}{\tau_en_{k,t}}\right)^2 +\frac{V\tau_uN_0\ln(2)}{h_k^u}\exp\left(\frac{R_{k,t}^u\ln(2)}{B_{k,t}}\right)\nonumber\\
    &+(Q_{k,t}^r-Q_{k,t}^u+\epsilon_zZ_k\varphi(n_{k,t})-\epsilon_z Z_kH_k^{\textrm{th}})\frac{\tau_u}{n_k}=0,
\end{align}
and bounding the solution according to constraint $(b)$ of \eqref{problem_edge_inference}. Note that \eqref{nonlineq} can be efficiently solved via a bisection method. Finally, once \eqref{nonlineq} is solved for each $n_{k,t}\in\mathcal{N}_k\setminus\{n_{k,t}=0\}$, the global optimal data rate is obtained as $R_{k,t}^{*}=\underset{n_{k,t}}{\arg\min}\;O_{r,t}$, also considering the case with $n_{k,t}=0$), with $O_{r,t}$ is the objective function of \eqref{radio_resource}.
\subsection{Mobile Edge Host CPU scheduling}
The second sub-problem, to optimize the MEH's CPU scheduling, can be formulated as follows:
\begin{align}\label{slot_opt_freq}
&\hspace{-.3 cm}\underset{\{f_k^r\}_k} \max \quad  \displaystyle \sum_{k=1}^K \tau f_{k,t}^r\frac{Q_{k,t}^r}{J_k^c}\;\;\displaystyle \\
&\hbox{{\rm s.t.}}\;(a)\: f_{k,t}^r\geq 0,\,\forall t,\quad (b)\:f_{k,t}^r\leq\min\left(f_{r}^{\max},\frac{Q_k^rJ_k^c}{\tau}\right)\nonumber\\
&(c)\: \sum\nolimits_{k=1}^Kf_{k,t}^r\leq f_r^{\max},\,\forall t.\nonumber
\end{align}
In \eqref{slot_opt_freq}, to improve efficiency w.r.t problem \eqref{problem_edge_inference}, we added, without loss of generality, constraint $(b)$, to ensure that each user cannot be assigned a clock frequency higher than the one necessary to empty the queue.
Problem \eqref{slot_opt_freq} is linear, and its global optimal solution consists in assigning the whole available frequency to the user with the highest ratio $Q_{k,t}^r/J_k^c$. If this leaves available frequency, the left part is assigned to the subsequent users until draining the whole CPU power of the server $f_r^{\max}$, or serving all users \cite{Merluzzi2020URLLC}.
\vspace{0 cm}
\section{Numerical Results}
\begin{figure*}[htb!]
    \centering
    \subfloat[Average E2E delay vs. device energy]{
        \includegraphics[width=0.32\textwidth]{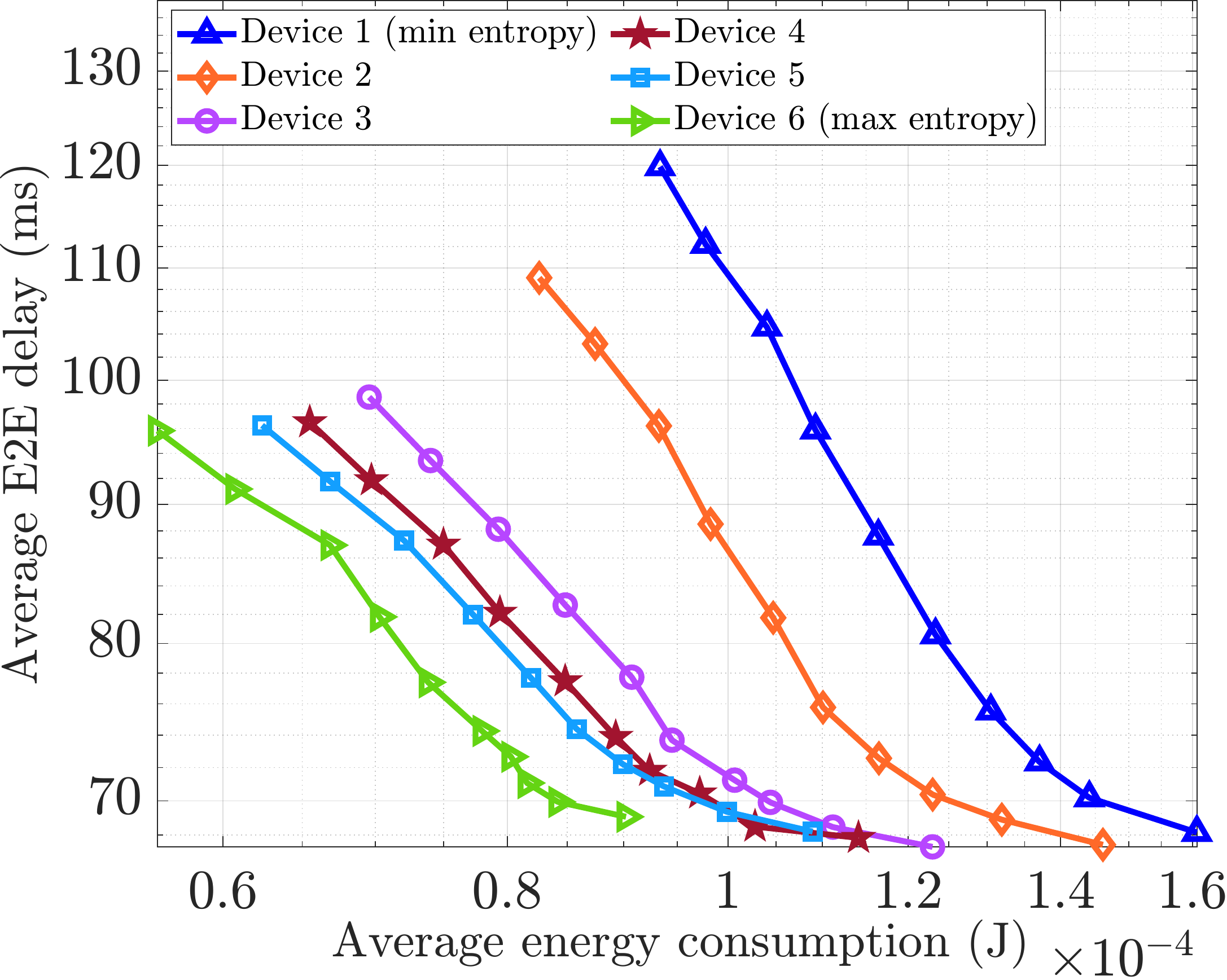}
        \label{fig:delay_vs_energy}
    }
    \subfloat[Average entropy vs. $t$]{
        \includegraphics[width=0.31\textwidth]{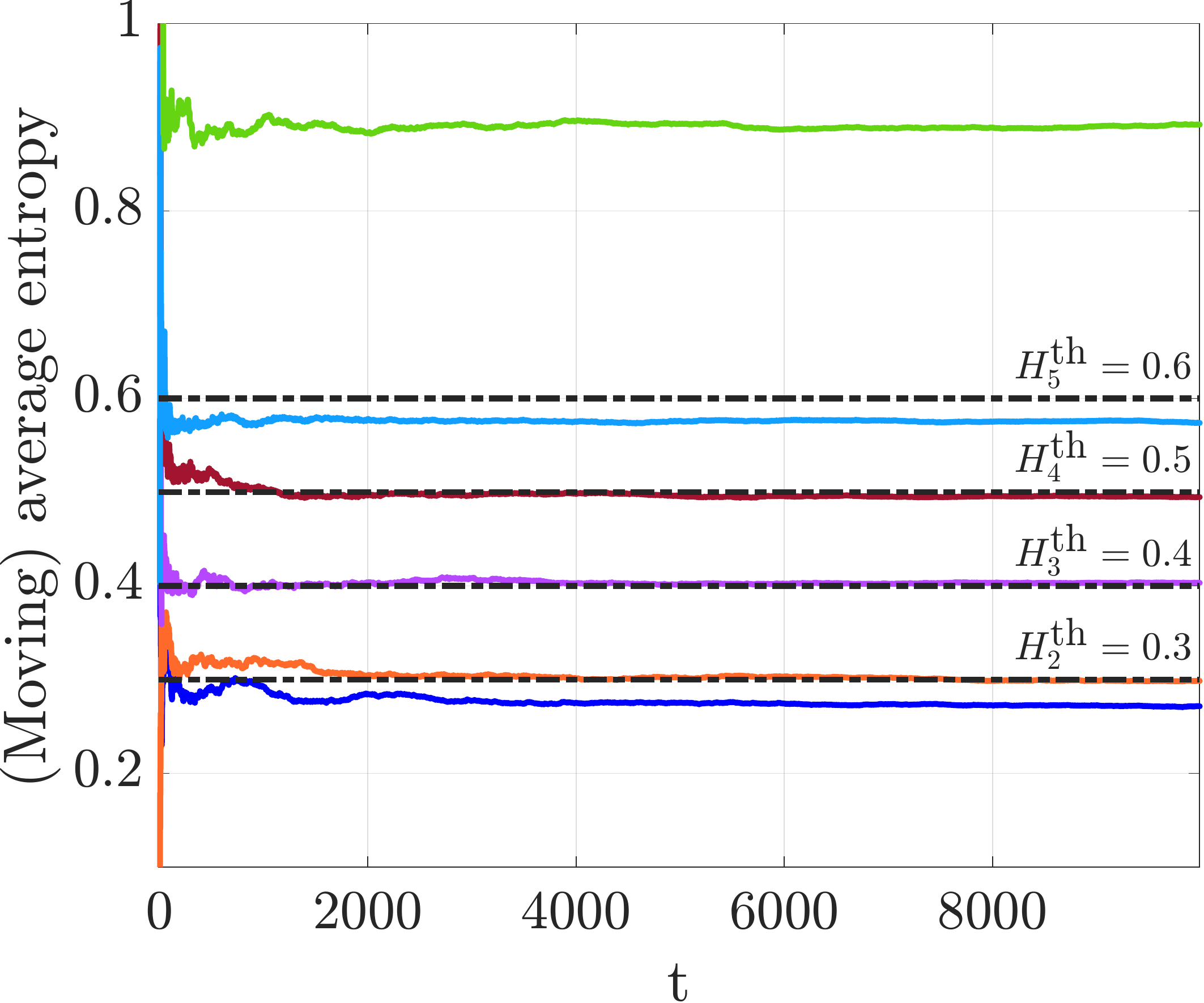}
        \label{fig:entropy_vs_t}
    }
    \subfloat[Correct classification rate vs. $t$]{
        \includegraphics[width=0.325\textwidth]{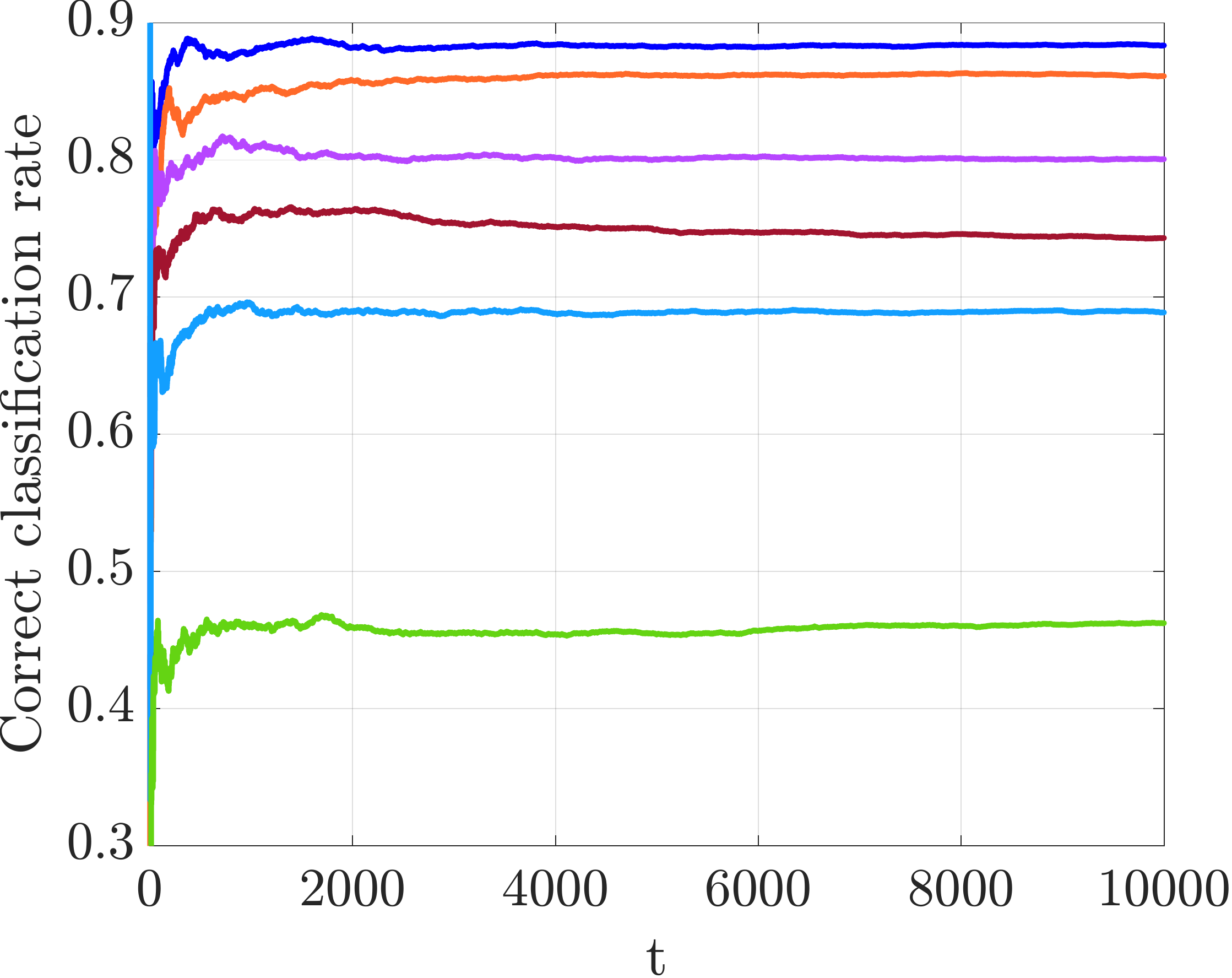}
        \label{fig:accuracy_vs_t}
    }
    \caption{Trade-off between average energy consumption, E2E delay, and inference reliability/accuracy}
    \label{fig:tradeoff}
    \vspace{0 cm}
\end{figure*}
In this section, we assess the performance of our algorithm. Let us briefly present the learning related setting, followed by the communication-computation scenario and parameters.\\
\underline{\textbf{\textit{Learning setting:}}} We recall that, since we focus on inference, we assume a pre-trained model, pre-uploaded in the MEH. For our numerical results, we consider the CIFAR-10 dataset\footnote{More results on other data sets are omitted due to the lack of space and left for an extension, but would come with similar conclusions (see Fig. \ref{fig:acc_entr_jpeg}).}. CIFAR-10 consists of 60000 32x32 color images divided in 10 classes, with 6000 images per class. The training set is made of 50000 instances, while the test set is made of 10000 instances. We exploit a Deep Convolutional Neural Network trained via usual backpropagation with an $l_2$-penalized Cross-entropy loss. The CNN architecture is made of six convolutional layers divided in three blocks with 32, 32 (end of the first block), 64, 64 (end of the second block), 128 and 128 (end of the third block) $3\times 3$ filters, respectively, with final flatten and dense layers; SAME padding, eLu non-linearities and Batch Normalization are applied after each convolutional layer, $2$ Max Pooling and Dropout is performed after each block and a final Softmax non-linearity is applied after the flatten and dense layers\footnote{https://appliedmachinelearning.blog/2018/03/24/achieving-90-accuracy-in-object-recognition-task-on-cifar-10-dataset-with-keras-convolutional-neural-networks/}.\\ 
\underline{\textbf{\textit{Communication and computation setting:}}} We consider a wireless AP at the center of a circle of radius $100$ m, endowed with an MEH with $f_r^{\max}=10$ GHz, and $6$ devices uniformly randomly located inside the circle, transmitting at maximum power $P_k^u=20$ dBm. Denoting by $f_c$ the carrier frequency in GHz, and by $d_k$ the distance in meters between device $k$ and the AP, the channel $h_{k,t}^u$ is generated with path loss $\textrm{PL}_k=33+25.50\log_{10}(d_k)+20\log_{10}(f_c)$, with a time varying Rayleigh fading with complex distribution $\mathcal{N}(0,1)$. The considered total available bandwidth is $100$ MHz, to be equally shared among all devices, while the noise power spectral density is $N_0=-174$ dBm/Hz, with a noise figure of $5$ dB at the receiver. We assume a slot duration $\tau=25$ ms, which is equally split among $\tau_e$ for encoding and $\tau_u$ for transmission. The effective switched capacitance of the processor is $\kappa_k=10^{-27}, \forall k$ (cf. \eqref{encoding_energy}), the number of CPU cycle to encode one pattern in set to $J_k^e=5\times10^5$ CPU cycles (cf. \eqref{uplink_data}), while the number of CPU cycles needed to classify one pattern is $J_k^c=10^7$ CPU cycles (cf. \eqref{num_computed}). We assume Poisson arrivals with parameter $\overline{A_k}=2,\forall k$. Each device has a different requirement in terms of $H_{k}^{\textrm{th}}$ (cf. \eqref{entropy_constraint}). In particular, two devices will serve as benchmarks for comparison, since they require, respectively, the highest and the lowest inference reliability. For the others, the requirement is $\mathbf{H}^{\textrm{th}}=[0.3,0.4,0.5,0.6]$. As surrogate function $\varphi(n_{k,t})$, we use a look up table that has been built on a validation set.\\
\underline{\textbf{\textit{Comments to the results:}}} Let us now comment the results in Fig. \ref{fig:tradeoff}, where we plot the energy-delay-reliability trade-off. In particular, in Fig. \ref{fig:delay_vs_energy}, we show the trade-off between E2E delay and energy consumption, obtained by increasing the single tuning parameter $V$ from right to left. The trade-off is different for each device, due to the different requirements in terms of inference reliability. As expected, all energies decrease as $V$ increases, with the cost of an increased E2E delay. Let us now focus on our two benchmarks, represented by the blue curve (highest reliability) and the green curve (lowest reliability). We can appreciate that the highest reliability ($H=0.27$, which translates into around $88$\% accuracy, as visible in Figs. \ref{fig:entropy_vs_t} and \ref{fig:accuracy_vs_t} that show the time average values of entropy and accuracy, respectively) is payed by a higher energy consumption (for a given E2E delay), due to the fact that more bits must be transmitted. On the other hand, achieving the "\textit{green solution}" (i.e., minimum energy consumption) is payed by a very low reliability ($H=0.88$, which translates into around $45$\% accuracy). However, among these two "extreme" cases, one can select different values of target entropy that the method is able to guarantee (see Fig. \ref{fig:entropy_vs_t} with the corresponding black dotted horizontal lines), while achieving lower energy consumption w.r.t. the highest accuracy case, but a much higher reliability and accuracy w.r.t. the minimum energy case. For instance, looking at the orange curve, we can conclude that a very small degradation in the accuracy (around $2$\%) is able to guarantee a non negligible gain in terms of energy-delay. Similar conclusions can be drawn for the other curves. More numerical results are omitted due to the lack of space, and are left for future extensions, for a deeper investigation of the achievable trade-offs in the context of edge inference.
\section{Conclusions}
In this paper, we proposed a method to jointly manage radio and computing resources to enable reliable and energy-efficient edge classification in dynamic scenarios. We efficiently solved a long-term problem through stochastic optimization tools, proposing a per-slot optimization (with quasi-closed form solutions), without requiring any apriori knowledge of the statistics of the involved random variables (i.e., wireless channels and data arrivals). To assess the performance of our method, we focused on a CNN based image classification, in which new images are adaptively encoded through different JPEG compression factors, to explore the fundamental trade-off between energy, delay, and accuracy.

\bibliographystyle{IEEEtran}
\bibliography{Mattia}

\end{document}